\documentclass{aip-cp}

\usepackage[numbers]{natbib}
\usepackage{rotating}
\usepackage{graphicx}

\begin{document}

\title{Dynamical Spin Effects in the Pion}
\author[aff1]{Mohammad Ahmady}
\eaddress{mahmady.mta.ca}
\author[aff3]{Farrukh Chishtie}
\eaddress{fchishti@uwo.ca}
\author[aff1,aff2]{Ruben Sandapen\corref{cor1}}
\affil[aff1]{Department of Physics, Mount Allison University, Sackville, New Brunswick E4L 1E6, Canada}
\affil[aff2]{Department of Physics, Acadia University, Wolfville, Nova Scotia B4P 2R6, Canada}
\affil[aff3]{Theoretical Research Institute, Pakistan Academy of Sciences (TRIPAS), Islamabad 44000, Pakistan}
\corresp[cor1]{Corresponding author: ruben.sandapen@acadiau.ca}

\maketitle

\begin{abstract}
We take into account dynamical spin effects in the holographic light-front pion wavefunction in order to predict the pion radius, decay constant, the pion electromagnetic and photon-to-pion transition form factors. We report a striking  improvement in the description of all data.
\end{abstract}

\section{INTRODUCTION}
A remarkable breakthrough in hadronic physics during the last decade is the discovery by Brodsky and de T\'eramond \cite{deTeramond:2008ht,Brodsky:2006uqa,deTeramond:2005su,Brodsky:2007hb,Brodsky:2011xx,Brodsky:2011yv} of a higher dimensional gravity dual to semiclassical light-front QCD.  This gauge/gravity duality, referred to as light-front holography, is reviewed in Reference \cite{Brodsky:2014yha}.  Light-front holography provides analytical forms as first approximations for hadronic light-front wavefunctions. In the semiclassical approximation, where quark masses and quantum loops are neglected, the hadron light-front wavefunctions depend on the invariant mass $\mathcal{M}^2=(\sum_{i}^n k_i)^2$ of the $n$ constituents rather than on their individual momenta $k_i$. For the valence ($n=2$ for mesons) Fock state, the invariant mass of the $q\bar{q}$ pair is $\mathcal{M}_{q\bar{q}}^2=k_{\perp}^2/x(1-x)$ and the latter is the Fourier conjugate to the impact variable $\zeta^2=x(1-x) b^2$ where $b$ is the transverse separation the quark and antiquark. The valence meson light-front wavefunction can then be written in a factorized form, 
\begin{equation}
	\Psi(\zeta, x, \phi)= e^{iL\phi} \mathcal{X}(x) \frac{\phi (\zeta)}{\sqrt{2 \pi \zeta}} \;,
\label{mesonwf}
\end{equation}
with the transverse mode satisfying a Schr\"odinger-like wave equation: 
\begin{equation}
 	\left(-\frac{d^2}{d\zeta^2} - \frac{1-4L^2}{4\zeta^2} + U(\zeta) \right) \phi(\zeta) = M^2 \phi(\zeta) \;,
 	\label{hSE}
 \end{equation}
where all the interaction terms and the effects of higher Fock states on the valence state are hidden in the confinement potential $U(\zeta)$.  This equation is said to be holographic because it maps (with $\zeta \to z_5$ where $z_5$ is the fifth dimension) onto the wave equation for the propagation of string modes in a $5$-dimensional anti-de Sitter space ($\mbox{AdS}_5$). In this AdS/QCD duality, the confining potential in physical spacetime is driven by the deformation of the pure $\mbox{AdS}_5$ geometry. Specifically, the potential is given by
 \begin{equation}
 	U(z_5, J)= \frac{1}{2} \varphi^{\prime\prime}(z_5) + \frac{1}{4} \varphi^{\prime}(z_5)^2 + \left(\frac{2J-3}{4 z_5} \right)\varphi^{\prime} (z_5) \;,
 \end{equation}
 where $\varphi(z_5)$ is the dilaton field which breaks  the maximal symmetry of $\mbox{AdS}_5$. A quadratic dilaton profile, $\varphi(z_5)=\kappa^2 z_5^2$, results in a light-front harmonic oscillator potential in physical spacetime, i.e.
 \begin{equation}
 	U(\zeta,J)= \kappa^4 \zeta^2 + \kappa^2 (J-1) \;.
 	\label{harmonic-LF}
 \end{equation}
 Group theoretical arguments by Brodsky, Dosch and de T\'eramond \cite{Brodsky:2013ar} point towards the uniqueness of  the quadratic term in Equation (\ref{harmonic-LF}). With the confining potential specified, one can solve the holographic Schr\"odinger equation to obtain the meson mass spectrum:
 \begin{equation}
 	M^2= 4\kappa^2 \left(n+L +\frac{S}{2}\right)\;,
 	\label{mass-Regge}
 \end{equation}
  which predicts (as required for massless quarks) a massless pion and the corresponding (normalized)  eigenfunctions given by
 \begin{equation}
 	\phi_{nL}(\zeta)= \kappa^{1+L} \sqrt{\frac{2 n !}{(n+L)!}} \zeta^{1/2+L} \exp{\left(\frac{-\kappa^2 \zeta^2}{2}\right)} L_n^L(x^2 \zeta^2) \;.
 \label{phi-zeta}
 \end{equation}
 To completely specify the holographic meson wavefunction,  we need the analytic form of the longitudinal mode $\mathcal{X}(x)$.  This is obtained by matching the expressions for the pion electromagnetic (EM) or gravitational form factor in physical spacetime and in AdS space. Either matching consistently results in $\mathcal{X}(x)=\sqrt{x(1-x)}$ \cite{Brodsky:2014yha,Brodsky:2008pf}. The pion holographic light-front wavefunction can thus be written as
\begin{equation}  
\Psi^{\pi} (x,b) = \frac{\kappa}{\sqrt{\pi}} \sqrt{x (1-x)}  \exp{ \left[ -{ \kappa^2 x(1-x)b^2  \over 2} \right] } \;.
\label{hwf}
\end{equation}

\section{QUARK MASSES AND HELICITIES}
The above AdS/QCD mapping leading to Equation (\ref{hwf}) is exact only for massless quarks and furthermore it does not specify how the wavefunction depends on quark helicities. For phenomenology, we need to restore the dependence on both quark masses and helicities. The quark-mass dependence is restored using the prescription of Brodsky and de T\'eramond \cite{Brodsky:2014yha} leading to
\begin{equation}  
\Psi^{\pi} (x,b) = \mathcal{N} \sqrt{x (1-x)}  \exp{ \left[ -{ \kappa^2 x(1-x) b^2  \over 2} \right] }
\exp{ \left[ -{m_f^2 \over 2 \kappa^2 x(1-x) } \right]} \;,
\label{hwfm}
\end{equation}
where $\mathcal{N}$ is a normalization constant fixed by requiring that
 \begin{equation}
 	\int \mathrm{d}^2 \mathbf{b}~\mathrm{d} x~|\Psi^{\pi}(x,b)|^2 = 1 \;.
 	\label{norm}
 \end{equation}
 In previous studies \cite{Swarnkar:2015osa,Vega:2008te,Vega:2009zb,Branz:2010ub}, the helicity dependence is assumed to decouple from the dynamics, i.e. the helicity wavefunction carries no momentum dependence. For the pseudoscalar pion, this means that
\begin{equation}
\Psi^{\pi}(x,\mathbf{k}) \to	 \Psi^{\pi}_{h,\bar{h}}(x, \mathbf{k}) =  \frac{1}{\sqrt{2}}~\bar{h}~ \delta_{-h, \bar{h}} \Psi^{\pi}(x,\mathbf{k})
\label{Spin-space}	\;,
\end{equation}
where $\Psi^{\pi}(x,\mathbf{k})$ is the two-dimensional Fourier transform of $\Psi^{\pi}(x,\mathbf{b})$.

Here, we assume a dynamical helicity dependence:
\begin{equation}
\Psi^{\pi}(x,\mathbf{k}) \to	 \Psi^{\pi}_{h\bar{h}}(x, \mathbf{k}) = S_{h\bar{h}}(x, \mathbf{k})  \Psi^{\pi}(x,\mathbf{k}) \;,
\label{Spin-space-dynamical}	
\end{equation}
where for the pseudoscalar pion,
\begin{equation}
	S^{\pi}_{h,\bar{h}}(x, \mathbf{k}) =\frac{\bar{v}_{\bar{h}}((1-x)P^+,-\mathbf{k})}{\sqrt{1-x}} \left[ (\alpha (P \cdot \gamma) +  M_{\pi} )\gamma^5 \right] \frac{u_{h}(xP^+,\mathbf{k})}{\sqrt{x}} 
	\label{spinwf}
\end{equation}
with $\alpha$ being an arbitrary complex constant. Note that for a vector meson like the $\rho$, choosing the spin structure $(\epsilon \cdot \gamma)M_{\rho}$ (where $\epsilon$ is the polarization vector of the $\rho$) in the square brackets of Equation (\ref{spinwf})  leads to impressive agreement with the HERA data on diffractive $\rho$ electroproduction  \cite{Forshaw:2012im}. In this contribution, we restrict ourselves to the special case where $\alpha=0$ in Equation (\ref{spinwf}). Our resulting spin-improved holographic wavefunction is therefore \cite{Ahmady:2016ufq}
\begin{equation}
\Psi^{\pi}_{h,\bar{h}}(x, \mathbf{k})= \mathcal{N}
M_{\pi}\bigg[\bar{h} \delta_{-h, \bar{h}} m_f  + \delta_{h, \bar{h}}  ke^{ih \theta_k}  \bigg] \frac{\Psi^{\pi} (x,\mathbf{k})}{x(1-x)} 
\label{spin-improved-k-space}	
\end{equation}
and the normalization constant $\mathcal{N}$ is now fixed using 
 \begin{equation}
 	\sum_{h,\bar{h}} \int \mathrm{d}^2 \mathbf{b} ~ \mathrm{d} x ~ |\Psi^{\pi}_{h,\bar{h}}(x,\mathbf{b})|^2 = 1 \;.
 	\label{norm-spin}
 \end{equation}

\section{COMPARISON WITH DATA}
It remains to specify the numerical value of the AdS/QCD scale $\kappa$ and the quark mass $m_{u/d}$ in order to predict observables. Previous work \cite{Brodsky:2014yha,Forshaw:2012im,Ahmady:2016ujw} has shown that $\kappa \approx 0.5$ GeV for light mesons. Specifically, we shall take the most recent universal value $\kappa=523$ MeV which fits the meson/baryon Regge slopes \cite{Brodsky:2016rvj} and accurately predicts $\Lambda^{{\mbox{\tiny{MS}}}}_{\mbox{\tiny{QCD}}}$ \cite{Deur:2016opc}. We shall predict the pion EM form factor, the pion radius and decay constant and the photon-to-pion transition form factor.  For the quark mass, we choose $m_{u/d}=330$ MeV, i.e. a constituent quark mass.

The EM form factor is given by
\begin{equation}
	F_{\pi}(Q^2)= 2 \pi \int \mathrm{d} x ~\mathrm{d} b ~ b ~ J_{0}[(1-x)  b Q] ~ |\Psi^{\pi}(x,\textbf{b})|^2 \;,
\label{DYW}
\end{equation}
where $|\Psi^{\pi}(x,\textbf{b})|^2$ is the pion light-front wavefunction squared and summed over all helicities. The pion radius can then be deduced from the slope of the EM form factor at $Q^2=0$:
\begin{equation}
	\langle r_{\pi}^2 \rangle = -\frac{6}{F_{\pi}(0)} \left. \frac{\mathrm{d} F_{\pi}}{\mathrm{d} Q^2} \right|_{Q^2=0} \;.
\end{equation}
The photon-to-pion transition form factor can be computed using 
\begin{equation}
	F_{\gamma \pi} (Q^2)= \frac{\sqrt{2}}{3} f_{\pi} \int_0^1 \mathrm{d} x \frac{\varphi_{\pi}(x,xQ)}{Q^2 x} \;,
	\label{TFF}
\end{equation}
where the twist-2 Distribution Amplitude $\varphi_{\pi}(x,\mu)$ and the decay constant $f_{\pi}$ can both be calculated using the holographic pion light-front wavefunction. Using our spin-improved holographic light-front wavefunction, we find that~\cite{Ahmady:2016ufq}
\begin{equation}
	f_{\pi}= 2 \sqrt{\frac{N_c}{\pi}} m_f M_{\pi}  \left.\int dx   \frac{\Psi^{\pi} (x,b)}{x(1-x)}\right|_{b=0} 
	\label{fpi-spin}
\end{equation}
and 
\begin{equation}
	f_{\pi} \varphi_{\pi}(x,\mu)= 2 m_f M_{\pi} \sqrt{\frac{N_c}{\pi}} \int \mathrm{d} b ~J_{0}(\mu b)~b   \frac{\Psi^{\pi} (x,b)}{x(1-x)} \;.
	\label{DA}
\end{equation}
We note that our predicted DA bears similarities with the platykurtic pion DA of Reference \cite{Stefanis:2014nla}.
 
\begin{figure}[h]
\label{fig:FF}
 \includegraphics[width=200pt]{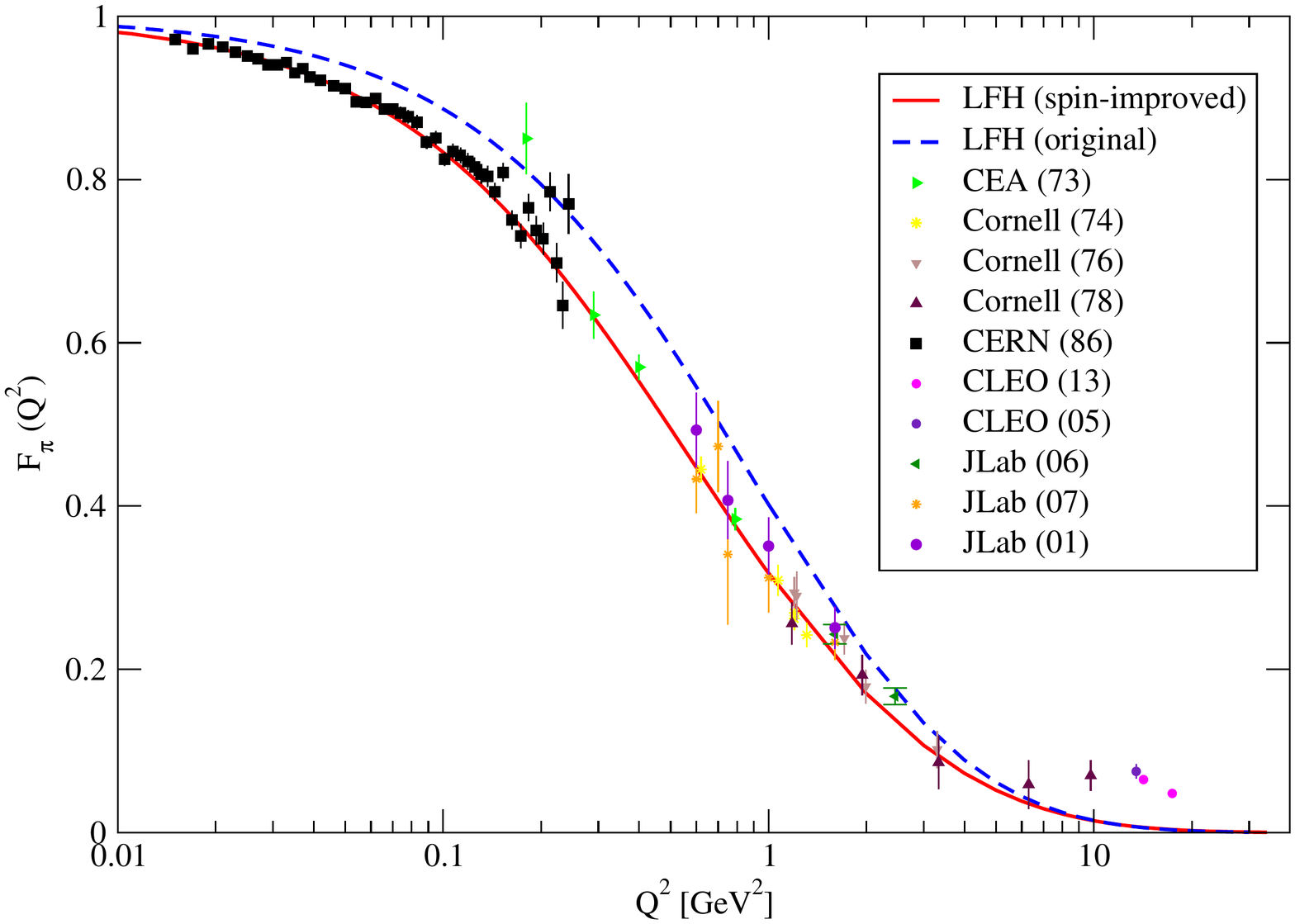} \includegraphics[width=200pt]{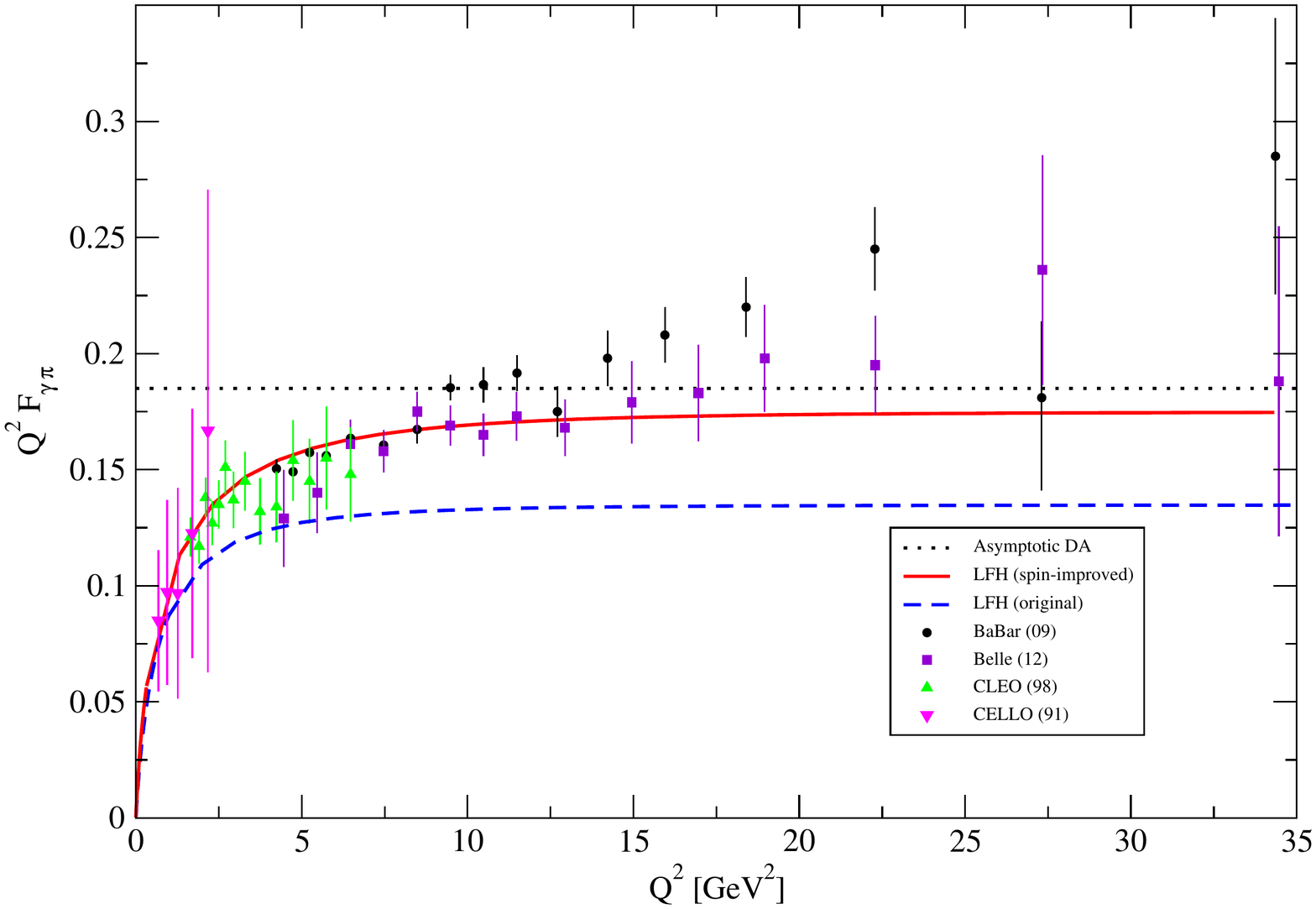}  
 \caption{Our predictions for the pion EM form factor (left) and the photon-to-pion transition form factor (right) using the original holographic wavefunction given by Equation (\ref{Spin-space}) (dashed-blue curve) and our spin-improved holographic wavefunction  given by Equation (\ref{spin-improved-k-space}) (continuous-red curve). The references for the data can be found in \cite{Ahmady:2016ufq}.}
\end{figure}

Our predictions for the pion radius and decay constant are shown in Table \ref{tab:fpi-R} and our results for the form factors are shown in Figure \ref{fig:FF}. The improvement in describing the data is remarkable for all observables.

\begin{table}[h]
\caption{Our predictions for the pion decay constant and charge radius using $\kappa=523$ MeV and $m_{u/d}=330$ MeV.}
\label{tab:fpi-R}
\tabcolsep7pt\begin{tabular}{lccc}
\hline
  & \tch{1}{c}{b}{Original}  & \tch{1}{c}{b}{Spin-improved}  & \tch{1}{c}{b}{Experiment}    \\
\hline
$f_{\pi}$  & 114 MeV & 135 MeV &  130.4 $\pm$ 0.04 $\pm$ 0.2  MeV\\
\hline
$\sqrt{\langle r_{\pi}^2 \rangle}$  & 0.544 fm & 0.683 fm & 0.672 $\pm$ 0.008 fm\\
\hline
\end{tabular}
\end{table}

\section{CONCLUSIONS}
We have shown that dynamical spin effects in the pion leads to a remarkable improvement in the description of pion observables. Our predictions are generated with the universal AdS/QCD scale, $\kappa= 523$ MeV, together with a constituent quark mass, $m_{u/d}= 330$ MeV. Our findings suggest that dynamical spin effects are important in the pion and it remains to investigate such effects with more general spin structures, i.e. with $\alpha \ne 0$ in Equation (\ref{spinwf}).


\section{ACKNOWLEDGMENTS}
This research is supported by a Team Discovery Grant from the Natural Sciences and Engineering Research Council (NSERC) of Canada. R.S. thanks the organizers of Diffraction 2016 for a very successful workshop.

\nocite{*}
\bibliographystyle{aipnum-cp}%
\bibliography{sandapen}%

\end{document}